\RequirePackage{fix-cm}
\documentclass{svjour3}                     
\smartqed  
\usepackage{graphicx}
\usepackage[cmex10]{amsmath}
\usepackage{amssymb}
\newcommand{\sgn}{\mathop{\mathrm{sgn}}}
\begin{document}

\title{Concrete incompleteness of Bell's correlation formula.
}


\author{Han Geurdes
}


\institute{J.F. Geurdes \at
              Tel.: +123-45-678910\\
              Fax: +123-45-678910\\
              \email{han.geurdes@gmail.com}           
}

\date{Received: Januari 5 2015 / Accepted: date}

\maketitle

\begin{abstract}
For a subset of 2 dimensional unit parameter vectors, Bell's correlation formula reproduces the quantum correlation. It is demonstrated with an example that the derivation of inequalities exclude beforehand certain local hidden variable models. In addition, the heuristic aspects of Bell's methodology are inspected.
\end{abstract}

\section{Introduction}
Bell's correlation function is a general expression for entanglement corelation in measurement parameters where hidden variables are employed in the explanation for $A\leftarrow S \rightarrow B$ type of experiments \cite{1}
\begin{equation}\label{B1}
E(a,b)=\int_{\lambda \in \Lambda} \rho(\lambda)A(a,\lambda)B(b,\lambda)d\lambda
\end{equation}
We consider all restrictions on local hidden models valid and known. The additional local parameters explanation of entanglement was initiated by Einstein \cite{2}.

\section{The  structure.}
Before presenting the details of the (partial) model, the model intended to test the strength of conclusions from inequalities. Such test can be defended by noting that there is no direct way of disproving whether locality principles are operative in entanglement. Use is made of  mathematical statistics. Excluding inequalities need to be fair in view of probability and in view of completeness. 

In $R^2$ we take e.g. $a^{T}=\left(\cos(\phi_a),\sin(\phi_a)\right)$. The fact that we focus on 2 dim parameter space is no real restriction to the criticism we will formulate. We may subsequently introduce $\phi$ in $[0,2\pi]$ that determine the setting vector. The following simple integral expression obtains:
\begin{equation}\label{B3}
a^{T}\cdot b = \int_0^{2\pi} d\phi_1 \delta(\phi_1-\phi_a)\int_0^{2\pi} d\phi_2 \delta(\phi_2-\phi_b)\cos(\phi_1 - \phi_2)
\end{equation}
The density functions for $\phi_j$, with $j=1,2$ are Dirac delta's and (\ref{B3}) is an expression of $a^{T}\cdot b = \cos(\phi_a - \phi_b)$. Note that, $\delta(\phi-\phi^{\prime})\geq 0$ and $\int_{0}^{2\pi}d\phi \,\delta(\phi-\phi^{\prime})=1$, for arbitrary $\phi^{\prime} \in (0,2\pi)$. The next question will be to find $f:[0,2\pi] \rightarrow [-1,1]$ and $g:[0,2\pi] \rightarrow [-1,1]$ such that $f(\phi_1)g(\phi_2)=\cos(\phi_1-\phi_2)$. 

Let us start our search for the $f$ and $g$ with the following.
\begin{eqnarray}\label{B6}
\begin{array}{ll}
\chi_1=\frac{\nu_1}{4} \pi + \frac{1}{2}\theta_1\\
\chi_2=\frac{\nu_2 }{4}\pi - \frac{1}{2}\theta_2
\end{array}
\end{eqnarray}
Here $\theta_1=\theta_1(\phi_1)$ and $\theta_2=\theta_2(\phi_2)$ with $\theta_j=\phi_j \pm  n\delta\phi$ and $j=1,2$ here. Note  $n$ a positive integer, including zero and $\delta\theta \geq 0$. If $\nu_1 +\nu_2=1$ and $\theta_1-\theta_2=\phi_1-\phi_2$ then $\psi=\chi_1+\chi_2$. The $\nu$ and $\theta$ are "form parameters"  of the measurement functions. 

Subsequently, $\sin(\psi)=\sin(\chi_1+\chi_2)$ and $\cos(\psi)=\cos(\chi_1+\chi_2)$. Using basic trigonometric rules it follows $\sin(\psi)=\sin(\chi_1)\cos(\chi_2)+\cos(\chi_1)\sin(\chi_2)$ \\ and $\cos(\psi)=\cos(\chi_1)\cos(\chi_2)-\sin(\chi_1)\sin(\chi_2)$. So,  
\begin{eqnarray}\label{B6a}
\begin{array}{ll}
\sin(\psi)\cos(\psi)=\\\\
\sin(\chi_1)\cos(\chi_1)\cos^2(\chi_2)-\sin^2(\chi_1)\sin(\chi_2)\cos(\chi_2)\\\\
+\cos^2(\chi_1)\sin(\chi_2)\cos(\chi_2) - \sin(\chi_1)\cos(\chi_1)\sin^2(\chi_2)
\end{array}
\end{eqnarray}
 it is possible to derive, using $\sin(2\chi) = 2\sin(\chi)\cos(\chi)$ that
\begin{eqnarray}\label{B7}
\begin{array}{ll}
\sin(\psi)\cos(\psi)=\frac{1}{2}\cos(2\chi_1)\sin(2\chi_2)+\frac{1}{2}\sin(2\chi_1)\cos(2\chi_2)
\end{array}
\end{eqnarray}
On the left hand of (\ref{B7}) we have terms that contain $\phi_1-\phi_2$. On the right hand of (\ref{B7}) a sum of products of terms with $\theta_1(\phi_1)$ in $\chi_1$ and $\theta_2(\phi_2)$ in $\chi_2$.  Let us introduce the hidden variable $\tau$ with uniform density on the real interval $[0,1]$. Hence, $\int_0^1 \rho(\tau) d\tau= \int_0^1 d\tau = 1$. Furthermore, let us define $I_1=[0,\frac{1}{2}] $ and $I_2=(\frac{1}{2},1]$. The indicator function $\iota(\tau \in I)$ is given by
\begin{eqnarray}\label{B8}
\iota(\tau \in I)=\{
\begin{array}{ll}
1, ~ \tau \in I\\
0,~ \tau \notin I
\end{array}
\end{eqnarray}
The following measurement functions can be defined. We have introduced the $\alpha$ and $\beta$ factors in the definitions for clarity. So,
\begin{eqnarray}\label{B9}
\begin{array}{ll}
A_0(\chi_1,\tau,\chi_3)=\iota(\tau \in I_1) \alpha \cos(2\chi_1) \sgn(\chi_3) \\\\
\hspace{0.8in}+\iota(\tau \in I_2)\alpha^{\prime}\sin(2\chi_1)\sgn(\chi_3)
\end{array}
\end{eqnarray}
together with
\begin{eqnarray}\label{B10}
\begin{array}{ll}
B_0(\chi_2,\tau,\chi_3)=-\iota(\tau \in I_1) \beta \sin(2\chi_2)\sgn(\chi_3) \\\\
\hspace{0.85in} - \iota(\tau \in I_2)\beta^{\prime}\cos(2\chi_2)\sgn(\chi_3)
\end{array}
\end{eqnarray}
Note, $\gamma=\alpha\beta=\alpha^{\prime}\beta^{\prime}=2$ and $\chi_3 \sim N(0,1)$ such that, $\sgn(\chi_3) \in \{-1,1\}$. Because of symmetry,
\[
\int_{-\infty}^{\infty}d\chi_3 \frac{e^{-\chi_3^2/2}}{\sqrt{2\pi}}\sgn(\chi_3)=0 
\]
and
\[
\int_{-\infty}^{\infty}d\chi_3 \frac{e^{-\chi_3^2/2}}{\sqrt{2\pi}}\{\sgn(\chi_3)\}^2=1 
\]
We have 
$
\mathcal{A}_{01}^{\not\equiv 0}= \alpha \cos(2\chi_1)
$
and  
$
\mathcal{A}_{02}^{\not\equiv 0}=\alpha^{\prime} \sin(2\chi_1)
$ 
for $\chi_1 \propto \phi_1$ in $A_0$. Similarly,
$
\mathcal{B}_{01}^{\not\equiv 0}=-\beta \sin(2\chi_2)
$
and 
$
 \mathcal{B}_{02}^{\not\equiv 0}=-\beta^{\prime}\cos(2\chi_2)
$
for $\chi_2 \propto \phi_2$. If we suppose that it is possible for certain $(\phi_j,\nu_j,\theta_j)$ to have both $\mathcal{A}_{0j}$ and $\mathcal{B}_{0j}$ at the same instance of $(\phi_1,\phi_2)$  in $[-1,1]$ for $j=1,2$ then the integration over $\tau$ in the interval $[0,1]$ will show, $A_0=A_0(\chi_1,\tau)$ and $B_0=B_0(\chi_2,\tau)$
\[
\int_0^1 d\tau  A_0B_0= -\frac{\gamma}{2} \left[\cos(2\chi_1)\sin(2\chi_2)+\sin(2\chi_1)\cos(2\chi_2)\right]
\]
This is true because
\begin{eqnarray}
\begin{array}{ll}
\int_0^1 d\tau \{\iota(\tau \in I_1)\}^2=\int_0^{1/2}d t= \frac{1}{2} \\\\
\int_0^1 d\tau \{\iota(\tau \in I_2)\}^2=\int_{1/2}^1 d t= 1-\frac{1}{2}= \frac{1}{2} \\\\
\int_0^1 d\tau \, \iota(\tau \in I_1) \iota(\tau \in I_2)=0
\end{array}
\end{eqnarray}
We already demonstrated that it is possible to have $|A|\leq 1$ and $|B|\leq 1$. So, $\gamma=2$ and $2\sin(\psi)\cos(\psi) = \sin(2\psi)=\cos(\phi_1-\phi_2)$ from (\ref{B3}) it follows that 
\begin{eqnarray}\label{B15}
\begin{array}{ll}
a^{T}\cdot b = -\int_0^{2\pi} d\phi_1 \delta(\phi_1-\phi_a)\int_0^{2\pi} d\phi_2 \delta(\phi_2-\phi_b)\\
~~~~~~~\times \int_0^1 dt A_0(\chi_1(\phi_1),\tau)B_0(\chi_2(\phi_2),\tau)
\end{array}
\end{eqnarray}
or more simpler, $\rho(\tau)=1$ for $\tau \in [0,1]$,
\begin{equation}\label{B16}
-\cos(\phi_a-\phi_b)=\int_0^1  \rho(\tau) \tilde{A}_0(\phi_a,\tau)\tilde{B}_0(\phi_b,\tau) \, d\tau
\end{equation}
$|\tilde{A}_0|\leq 1$ and $|\tilde{B}_0|\leq 1$ such that $\tilde{A}_0(\phi_a,\tau)=A_0(\chi_1(\phi_a),\tau)$ and $\tilde{B}_0(\phi_b,\tau)=B_0(\chi_2(\phi_b),\tau)$. So, (\ref{B16}) is a general correlation like in  (\ref{B1}). The transformation of $\phi_a$ to $\chi_1$ which is defined by $\nu_1$ and $\theta_1$. Similarly for $\phi_b$ by the form in (\ref{B6}). The reader notes that (\ref{B16}) employs separation of the setting variables $\phi_a$ and $\phi_b$ in their transformed formats $\chi_1$ and $\chi_2$. This is what is required. We must, however, demonstrate that $\mathcal{A}_{0j} \in [-1,1]$ and simultaneously $\mathcal{B}_{0j} \in [-1,1]$, with $j=1,2$ is reallypossible.  In the table below it is numerically demonstrated that for some $(\phi_a,\phi_b)$ pairs, $\alpha, \beta, \alpha^{\prime}, \beta^{\prime}, \nu_1,\nu_2,\theta_1$ and $\theta_2$ can be found such that  (\ref{B16}) is a valid expression. I.e. $A_0=\mathcal{A}_{01}+\mathcal{A}_{02}$ and $B_0=\mathcal{B}_{01}+\mathcal{B}_{02}$, thereby observing the term definitions in (\ref{B9}) and (\ref{B10}). 
\begin{table}[ht]
\caption{$\phi_a$ and $\phi_b$ values (radians), $\lambda$ and $\theta$ for $\mathcal{A} \in [-1,1]$ and $\mathcal{B} \in [-1,1]$, $\alpha \beta=\alpha^{\prime}\beta^{\prime}=2$. Note $\nu_2=1-\nu_1$.  The values are presented in two decimals approximation. }
\centering
\begin{tabular}{c c c c c c c c c c}
\hline
$\phi_a$ & $\phi_b$ & $\nu_1$ & $\theta_1$ & $\theta_2$ & $\alpha$ & $\beta$ & $\alpha^{\prime}$ & $\beta^{\prime}$ \\
\hline
2.87 & 1.80 & 0.90 &  2.86 & 1.80 &  2.00 & 1.00 & 1.10 & 1.82 \\
4.96 & 5.86 & 0.16 &  4.96 & 5.85 &  2.00 & 1.00 & 1.14 & 1.76 \\
1.16 & 5.66 & 0.61 & 1.16 & 5.66 &  1.92 & 1.04 & 1.17 & 1.71 \\
\hline
\end{tabular}
\label{TB1}
\end{table}
\begin{table}[ht]
\caption{$\phi_a$ and $\phi_b$ values (radians) that allow  $\mathcal{A} \in [-1,1]$ and $\mathcal{B} \in [-1,1]$, $\alpha \beta=\alpha^{\prime}\beta^{\prime}=2$. The $\mathcal{A}$ and $\mathcal{B}$ derive from table \ref{TB1} and equations (\ref{B9}) and (\ref{B10}) and are presented in two decimals approximation. }
\centering
\begin{tabular}{c c c c c c c c c}
\hline
$\phi_a$ & $\phi_b$ & $\mathcal{A}_{01}^{\not\equiv 0}$ & $\mathcal{A}_{02}^{\not\equiv 0}$ & $\mathcal{B}_{0,1}^{\not\equiv 0}$ & $\mathcal{B}_{0,2}^{\not\equiv 0}$\\
\hline
2.87 & 1.80 & -0.83 & -0.99 & -0.99 & -0.12\\
4.96 & 5.86 & 0.95 & -0.99 & 0.98 & -0.31 \\
1.16 & 5.66 & -0.99 & 0.99 & 0.98 & 0.56 \\
\hline
\end{tabular}
\label{TB2}
\end{table}
For ease of presentation, 
\[
\mathcal{A}_{0j}=\iota(\tau \in I_j) \{\delta_{1,j}\alpha \cos(2\chi_1)+ \delta_{2,j}\alpha^{\prime} \sin(2\chi_1)\}\sgn(\chi_3)
\]
and 
\[
\mathcal{B}_{0j}=-\iota (\tau \in I_j) \{\delta_{1,j}\beta \sin(2\chi_2)+ \delta_{2,j}\beta^{\prime} \cos(2\chi_2)\}\sgn(\chi_3)
\]
where $\delta_{i,j}$ Kronecker's delta, $I_1 \cap I_2 =\emptyset$, $\iota(...)$ an indicator function  and $j=1,2$. The resulting measurement functions can project in $\{-1,1\}$  when the zero indexed functions are transformed into
\begin{eqnarray}\label{ext1}
\begin{array}{ll}
\mathcal{A}_1=\sgn\{{A}_0-\lambda_A\} \\
\mathcal{B}_1=\sgn\{{B}_0-\lambda_B\}
\end{array}
\end{eqnarray}  
The $\sgn(x)=1$ for $x\geq 0$ and $\sgn(x)=-1$ for $x<0$. Moreover, $A_0=\mathcal{A}_{01}+\mathcal{A}_{02} \in [-1,1]$ and $B_0=\mathcal{B}_{01}+\mathcal{B}_{02} \in [-1,1]$, noting (\ref{B9}) and (\ref{B10}) and $\mathcal{A}_1\in \{-1,+1\}$ together with $\mathcal{B}_1\in \{-1,+1\}$. The hidden variables $\lambda_A$ and $\lambda_B$ are uniform distributed variables in $[-1,1]$ with $\rho_{\lambda_A}=\frac{1}{2}$ and $\rho_{\lambda_B}=\frac{1}{2}$. From table \ref{TB1} and \ref{TB2} we can deduce that {\em the construction} of the measurement functions is not local. This means that the construction of the functions must be performed before the start of a statistics experiment. Subsequently, we investigate whether the derivation of the Bell inequality exclude beforehand certain local models. Let us assume $E(a,b)=E(b,a)$ and introduce a short-hand notation for the integration in (\ref{B1}).  Note that if the $E(a,b)=E(b,a)$ is disallowed then local models to explain $-(a\cdot b)$ are disqualified beforehand. 
\subsection{Bell operations on a related measure}
The following elementary equality is derived from Bell's correlation formula in (\ref{B1}).
We start by defining $[ab]_{AA}$ 
\begin{equation}\label{Extr0a}
[ab]_{AA}=\int d\lambda \rho(\lambda)\mathcal{A}_1(a,\lambda)\mathcal{A}_1(b,\lambda)
\end{equation}
The expression in (\ref{Extr0a}) can be rewritten as
\begin{eqnarray}\label{Extr0b}
\begin{array}{ll}
[ab]_{AA}=\\\\
\hspace{0.02in}=\int d\lambda \rho(\lambda)\mathcal{A}_1(a,\lambda)\mathcal{A}_1(b,\lambda)\{1-\mathcal{A}_1(a,\lambda)\mathcal{B}_1(b,\lambda)\}\\\\
\hspace{0.02in}+\int d\lambda \rho(\lambda)\mathcal{A}_1(b,\lambda)\mathcal{B}_1(b,\lambda)
\end{array}
\end{eqnarray}
Now, $\{1-\mathcal{A}_1(a,\lambda)\mathcal{B}_1(b,\lambda)\} \geq 0$ and $\mathcal{A}_1(a,\lambda)\mathcal{A}_1(b,\lambda)\leq 1$. We then see: $[ab]_{AA}\leq 1-E(a,b)+E(b,b)$. 
The $E(b,b)=-1$ cannot theoretically be denied and so, 
 $[ab]_{AA} \leq -E(a,b)$.
The equation in (\ref{Extr0a}) can also be rewritten as
\begin{eqnarray}\label{Extr1b}
\begin{array}{ll}
[ab]_{AA}=\\\\
\hspace{0.05in}=\int d\lambda \rho(\lambda)\mathcal{A}_1(a,\lambda)\mathcal{A}_1(b,\lambda)\{1+\mathcal{A}_1(a,\lambda)\mathcal{B}_1(b,\lambda)\}\\\\
\hspace{0.05in}-\int d\lambda \rho(\lambda)\mathcal{A}_1(b,\lambda)\mathcal{B}_1(b,\lambda)
\end{array}
\end{eqnarray}
With, $\{1+\mathcal{A}_1(a,\lambda)\mathcal{B}_1(b,\lambda)\} \geq 0$ and $E(b,b)=-1$ gives with, $\mathcal{A}_1(a,\lambda)\mathcal{A}_1(b,\lambda)\geq -1$, hence, 
 $[ab]_{AA} \geq -E(a,b)$.
Hence, \[[ab]_{AA}=-E(a,b)\] with $1=[bb]_{AA}=-E(b,b)$. From (\ref{B9}) and (\ref{B10}) it follows $\mathcal{A}_1(b,\lambda) \neq -\mathcal{B}_1(b,\lambda)$. Note that in terms of the model we have 
\begin{eqnarray}\label{Extr5}
\begin{array}{ll}
[bb]_{AA}=
\int_0^1 d\tau \int_{-\infty}^{\infty}d\chi_3 \frac{e^{-\chi_3^2/2}}{\sqrt{2 \pi}}\int_{-1}^{+1}\frac{d\lambda_A}{2}\int_{-1}^{+1}\frac{d\lambda_B}{2} \\\\
\hspace{.45in} \times \{\sgn^2(A_0(\phi_b,\chi_3,\tau)-\lambda_A)\times 1\}
\end{array}
\end{eqnarray}
This result implies that $[bb]_{AA}=1$ which agrees with $E(b,b)=-1$ and $[bb]_{AA}=-E(b,b)$. Furthermore let us evaluate $[ab]_{AA}$ in terms of the model. We must evaluate ($\chi_3$ dependence implicit)
\begin{equation}\label{Extra}
I=\int_{-1}^{+1}\frac{d\lambda_A}{2}\sgn(A_0(\phi_a,\tau)-\lambda_A)\sgn(A_0(\phi_b,\tau)-\lambda_A)
\end{equation}
Suppose $A_0(\phi_a,\tau)\geq A_0(\phi_b,\tau)$.  Hence,
\begin{eqnarray}\label{Extrc}
\begin{array}{ll}
I=\int_{-1}^{A_0(\phi_a,\tau)}\frac{d\lambda_A}{2}\sgn(A_0(\phi_b,\tau)-\lambda_A)\\\\
\hspace{.2in} -\int_{A_0(\phi_a,\tau)}^1 \frac{d\lambda_A}{2} \sgn(A_0(\phi_b,\tau)-\lambda_A)
\end{array}
\end{eqnarray}
So, under $A_0(\phi_a,\tau)\geq A_0(\phi_b,\tau)$,
\begin{equation}\label{Extrd}
I=\int_{-1}^{A_0(\phi_b,\tau)}\frac{d\lambda_A}{2}-\int_{A_0(\phi_b,\tau)}^{A_0(\phi_a,\tau)}\frac{d\lambda_A}{2}+\int_{A_0(\phi_a,\tau)}^1\frac{d\lambda_A}{2}
\end{equation}
This implies $I=1-[A_0(\phi_a,\tau)-A_0(\phi_b,\tau)]$ when $A_0(\phi_a,\tau)\geq A_0(\phi_b,\tau)$. With $(\tau,\chi_3)$ dependence implicit,
\begin{eqnarray}\label{Extrb}
\begin{array}{ll}
[ab]_{AA}=1-\int_0^1d\tau \int_{-\infty}^{\infty} d\chi_3 \frac{e^{-\chi_3^2/2}}{\sqrt{2\pi}} \\\\
\hspace{.3in} \{ H(A_0(\phi_a)-A_0(\phi_b))[A_0(\phi_a)-A_0(\phi_b)]\\\\
\hspace{.3in} +H(A_0(\phi_b)-A_0(\phi_a))[A_0(\phi_b)-A_0(\phi_a)]\}
\end{array}
\end{eqnarray}
$H(x)=1$, $x\geq 0$ and $H(x)=0,\,x<0$. Hence, because $(H(y)-H(-y))y=|y|$ for real $y$, 
\[
[ab]_{AA}=1-\int_0^1d\tau \int_{-\infty}^{\infty} d\chi_3 \frac{e^{-\chi_3^2/2}}{\sqrt{2\pi}}|A_0(\phi_a)-A_0(\phi_b)|
\]
From $\mathcal{A}_{0j}$ we deduce that $\chi_3$ dependence drops off in the absolute sign operation. This produces,
\[
 [ab]_{AA}=1-\int_0^1d\tau |\iota(\tau\in I_1)\xi+\iota(\tau\in I_2)\eta|
\]
Here, $\xi=\mathcal{A}_{01}^{\not\equiv 0}(\phi_a)-\mathcal{A}_{01}^{\not\equiv 0}(\phi_b)$ and $\eta=\mathcal{A}_{02}^{\not\equiv 0}(\phi_a)-\mathcal{A}_{02}^{\not\equiv 0}(\phi_b)$. Hence,
\[
 [ab]_{AA}=1-\int_0^{1/2}d\tau |\xi|-\int_{1/2}^1 d\tau |\eta|
\]
For $0\leq \tau \leq 1/2$  we have $\iota(\tau \in I_1)=1$ and $\iota(\tau \in I_2)=0$. For, $1/2 < t \leq 1$, $\iota(\tau \in I_1)=0$ and $\iota(\tau \in I_2)=1$. Hence,
\begin{equation}\label{Extrd1}
 [ab]_{AA}=1-\left(\frac{|\xi|+|\eta|}{2}\right)
\end{equation}
Numerically we will see that violating instances of a Bell type inequality observe $-1\leq [ab]_{AA} \leq 1 $.

\subsection{ Bell operations for inequalities.}
From equation (\ref{B1}) we can, for $A\in \{-1,+1\}$ and $B\in \{-1,+1\}$, derive
\begin{eqnarray}\label{TMP1}
\begin{array}{ll}
|E(a,b)-E(c,b)| =\\\\ 
\hspace{.2in} =|\int d\lambda \rho(\lambda)\{A(\lambda,a)B(\lambda,b)-A(\lambda,c)B(\lambda,b)\}|
\end{array}
\end{eqnarray}
Similar to the derivation of CHSH \cite{2a} we add and subtract $A(\lambda,a)B(\lambda,b)A(\lambda,c)B(\lambda,a)$ to (\ref{TMP1}). This gives
\begin{eqnarray}\label{TMP2}
\begin{array}{ll}
|E(a,b)-E(c,b)| =\\\\ 
\hspace{.04in} =|\int d\lambda \rho(\lambda)A(\lambda,c)B(\lambda,b)\{-1-A(\lambda,a)B(\lambda,a)\} \\\\
\hspace{.05in} +\int d\lambda \rho(\lambda)\left(-A(\lambda,a)B(\lambda,b)\right)\{-1-A(\lambda,c)B(\lambda,a)\}|
\end{array}
\end{eqnarray}
From (\ref{Extr5}), $E(a,a)=-1$ and $|A(\lambda,c)B(\lambda,b)|\leq 1$ together with $|-A(\lambda,a)B(\lambda,b)|\leq 1$,  then it is found that 
$ S^{\prime}=|E(a,b)-E(c,b)| -E(c,a)\leq 1$
The point made here is that the to be tested inequality is obtained with the use of principles similar to the derivation of the CHSH \cite{2a}.

\subsection{Numerics of violation}
In table-\ref{TB3} a Bell type inequality for triples $(\phi_a,\phi_b,\phi_c)$
$ S=|E(a,b)- E(c,a)|-E(b,c) \leq 1 $ 
is tested against the model developed here. This computation is set up to answer possible objections which could claim that it is not so very extremely worrying to exclude local models that cannot violate a Bell-type inequality. We employed: $E(a,a)=-1$ and $E(a,b)=E(b,a)$ etc. In the table below examples of explicit computation are given. The double bar separates the instances. The first example is to show a violation $S>1$ for $(\phi_a,\phi_b),\,\, (\phi_b,\phi_c)$ and $(\phi_c,\phi_a)$. The second two examples also include data for $[ca]_{AA}=-E(c,a)$ computation.
\begin{table}[ht]
\caption{$\phi_1$ and $\phi_1$ values (radians) in $(\phi_a,\phi_b,\phi_c)$ with $S=|E(a,b)- E(c,a)|-E(b,c) > 1$, in 2 decimals approximation.}
\centering
\begin{tabular}{c c c c c c c c c}
\hline
$\phi_1$ & $\phi_2$ & $\chi_1$ & $\chi_2$ & $\mathcal{A}_{01}^{\not\equiv 0}$ & $\mathcal{A}_{02}^{\not\equiv 0}$ & $\mathcal{B}_{0,1}^{\not\equiv 0}$ & $\mathcal{B}_{0,2}^{\not\equiv 0}$\\
\hline
0.82 & 5.16 & 0.86 & -2.24 & -0.29 & 0.99 &  0.97 & -0.45  \\
5.16 & 5.33 & 2.71 & -2.01 & 0.99 & -0.99 &  0.98 & -0.99  \\
5.33 & 0.82 & 3.26 & -0.22 & 0.99 & 0.48 &  -0.84 & 0.90 \\
\hline
\hline
0.89 & 5.40 & 0.76 & -2.23 & 0.10 & 0.99 &  0.97 & -0.50  \\
0.89 & 5.23 & 0.83 & -2.22 & -0.18 & 0.99 &  0.96 & -0.56  \\
5.40 & 5.23 & 2.79 & -1.92 & 0.99 & -0.99 &  0.98 & -0.99 \\
5.23 & 0.89 & 2.94 & 0.02 & 0.98 & -0.78 &  0.06 & 0.99\\
\hline
\hline
1.47 & 5.04 & 1.06 & -2.06 & -0.99 & 0.99 &  0.87 & -0.94  \\
1.47 & 5.40 & 0.98 & -2.17 & -0.77 & 0.99 &  0.93 & -0.69  \\
5.04 & 5.40 & 2.67 & -2.07 & 0.99 & -0.99 &  0.99 & -0.89 \\
5.40 & 1.47 & 3.01 & -0.26 & 0.99 & -0.51 &  -0.97 & 0.87\\
\hline
\end{tabular}
\label{TB3}
\end{table}
Table-\ref{TB3} shows $(\phi_a,\phi_b), (\phi_b,\phi_c)$ and $(\phi_c,\phi_a)$  i.e. the $\chi_1(\phi_{\cdot})$, $\chi_2(\phi_{\cdot})$ and $S>1$. In $(\phi_a,\phi_b,\phi_c)=(0.82, 5.16,5.33)$, $a=(\cos(0.82),\sin(0.82))=(0.68,0.20)$, $b=(\cos(5.16), \sin(5.16))=(0.43,-0.90)$,\\ $c=(\cos(5.33), \sin(5.33))=(0.58,-0.82)$. Hence, $Eab=-0,11, Ebc=-0,99$ and $Eca=-0,23$, which implies, $|Eab-Eca|-Ebc \approx 1.12$. Hence, $\sigma=S-1=0.12$, furthermore the second instance in table-\ref{TB3}, $\sigma=1.36$ and the third, $\sigma=0.14$.  
\begin{table}[ht]
\caption{$\phi_a$ and $\phi_c$  (radians) $[ab]_{AA}=1-(|\xi|+|\eta|)/2$ based on (\ref{Extrd1}) and $E(a,c)=-\cos(\phi_a-\phi_c)$, $|\xi|=|\mathcal{A}_{01}^{\not\equiv 0}(\phi_a)-\mathcal{A}_{01}^{\not\equiv 0}(\phi_c)|$, $|\eta|=|\mathcal{A}_{02}^{\not\equiv 0}(\phi_a)-\mathcal{A}_{02}^{\not\equiv 0}(\phi_c)|$ and $\Delta=|\,|E(a,c)|-|[ac]_{AA}|\,|$. }
\centering
\begin{tabular}{c c c c c c c c}
\hline
&$\phi_1$ & $\phi_2$ & $\mathcal{A}_{01}^{\not\equiv 0}(\phi_1)$ & $\mathcal{A}_{02}^{\not\equiv 0}(\phi_1)$ & $[ac]_{AA}$ & $E(a,c)$ & $\Delta$\\
\hline
a & 1.47 & 5.40 & -0.77 & 0.99 & - & - \\
c & 5.40 & 1.47 & 0.99 & -0.51 & -0.630 & 0.705 & 0.08\\
\hline
\hline
a & 0.89 & 5.23 & -0.18 & 0.99 & - & - \\
c & 5.23 & 0.89 & 0.98 & -0.78 & -0.465 & 0.364 &  0.10\\
\hline
\end{tabular}
\label{TB5}
\end{table}
In order to obtain the figures in table-\ref{TB5} we use table-\ref{TB3} and take the rows representing the combinations $(0.89,5.23)$ for $\phi_a=0.89$ data and $(5.23,0.89)$ for $\phi_c=5.23$ data. Similar for $(\phi_a,\phi_c)=(1.47,5.40)$. Table - \ref{TB5} shows two instances where $[ac]_{AA}\neq -E(a,c)$ numerically ($\Delta \sim 0.05 \, - \, 0.10$). 

\section{Conclusion \& discussion}
In the paper it is argued that no-go experiments using Bell inequalities may exclude beforehand certain local hidden variable models that {\em can} show critical violating parameters. Such a model was presented and checked numerically. The results of the paper show that it is not obvious to see {\em all} models under certain inequality operations. A local model which covers a relevant part of the setting parameter space, operationally defined, $[0,2\pi]\times [0,2\pi]$, violates a Bell inequality. The $(\phi_a,\phi_b)$ we found belong to $\Phi \subset [0,2\pi]\times [0,2\pi]$ a real subset. Moreover, $\Phi \subset \Phi_{viol}$ with $\Phi_{viol}$ the totality of violating pairs.  Notwithstanding this restriction, if using the Bell operations one unknowingly suppress some local models beforehand, then experimental no-go conclusions are premature. The functions presented in the numerical study can also be the basis of a statistical experiment resembling \cite{3}. 
We have dealt with data-analysis of the $A\leftarrow S \rightarrow B $ experiment. If $E(a,b)=E(b,a)$ is denied for the model, then it is impossible to model $-(a\cdot b)$. In addition, it is clear that 
\[
\int_0^1 d\tau \int_{-\infty}^{\infty}d\chi_3 \frac{e^{-\chi_3^2/2}}{\sqrt{2\pi}} A_0(\phi_a,\chi_3,\tau)=0
\]  
together with a similar expression for $B_0(\phi_a,\chi_3, \tau)$. Hence, the model shows the proper average values. Note that we do not present a complete explanation but just aim to inspect Bell's formula and its use. The $[ab]_{AA}=-E(a,b)$ result suggests that one sided measurements by Alice give the (negative of) the entanglement correlation between measurements $A$, by Alice, and $B$, by Bob. This means that the mathematics of the inequalities entails that the entangled correlation can be obtained from local, one sided, measurements by Alice alone. The usual interpretation where Alice sets $a$ and Bob $b$ gives the operationalization of the entanglement correlation
\begin{equation}\label{Dis1}
E(a,b)=\frac{N(+,+|ab)+N(-,-|ab)-N(+,-|ab)-N(-,+|ab)}{N(+,+|ab)+N(-,-|ab)+N(+,-|ab)+N(-,+|ab)}
\end{equation}
This is based on countings $N(s_A(t),s_B(t)|a(t),b(t))$ with equal time $(s_A(t),s_B(t))\in \{-1,1\}^2$ measurements at Alice and Bob's site, afer the experiment has ended. If the wings of the experiment $A\leftarrow S \rightarrow B$, are of unequal length, e.g. Euclidean norm for place vectors $||A-S||_e<||S-B||_e$, then unequal times measurements $(s_A(t_0),s_B(t_1))$ with $t_0<t_1$ processed in $N(s_A(t_0),s_B(t_1)|a(t_0),b(t_1))$ , will provide similarly an approximate operationalization of the correlation
\begin{equation}\label{Dis1a} 
E(a,b)(t_0,t_1)=\int d\lambda \rho(\lambda) A(a,\lambda)(t_0) B(b,\lambda)(t_1). 
\end{equation}
$N(s_A(t_0),s_B(t_1)|a(t_0),b(t_1))$ can be employed in the approximation of the weighted integral. Hence, the question can be raised if   
\begin{equation}\label{Dis1b} 
[ab]_{AA}=\int d\lambda \rho(\lambda) A(a,\lambda)(t_0) A(b,\lambda)(t_1) 
\end{equation}
can be approximated using $N_{AA}(s_A(t_0),s_A(t_1)|a(t_0),b(t_1))$ in 
\begin{equation}\label{Dis2}
[ab]_{AA}=-\frac{N_{AA}(=|a(t_0),b(t_1))-N_{AA}(\neq |a(t_0),b(t_1))}{N_{AA}(=|a(t_0),b(t_1))+N_{AA}(\neq |a(t_0),b(t_1))} \end{equation} 
with  $N_{AA}(=|a(t_0),b(t_1))$ equal to \[N_{AA}(+,+|a(t_0),b(t_1))+N_{AA}(-,-|a(t_0),b(t_1))\] and $ N_{AA}(\neq|a(t_0),b(t_1))$ equal to \[N_{AA}(+,-|a(t_0),b(t_1))+N_{AA}(-,+|a(t_0),b(t_1)).\] 
If it is argued that one-sided A measurements, in $t_0<t_1$ order, are absolutely random then $[ab]_{AA}=-E(a,b)$ operationalized as in (\ref{Dis2}) cannot be valid. The principles that lead to Bell inequalities then also may lead to contradictions. If the approximation of $[ab]_{AA}$ using $N_{AA}(s_A(t_0),s_A(t_1)|a(t_0),b(t_1))$ as in (\ref{Dis2}) is contested then one may ask why an equivalent rule is considered valid for (\ref{Dis1a}). There appears, mathematically, little difference between (\ref{Dis1a}) and (\ref{Dis1b}). In (\ref{Dis1a}) there are presently no objections to the use of countings $N(s_A(t_0),s_B(t_1)|a(t_0),b(t_1))$. So what will prevent using $N_{AA}(s_A(t_0),s_A(t_1)|a(t_0),b(t_1))$ in (\ref{Dis1b}) to obtain $[ab]_{AA}$. In a completely separated $(\lambda_A,\lambda_B)$ model one cannot rely on the notion that similar to the evaluation of the AB integral, i.e. (\ref{Dis1a}) with(\ref{Dis1}), Bob simply is using an A-type function in the evaluation of $[ab]_{AA}$, (\ref{Dis1b}). But (\ref{Dis2}) is based on A side measurements and we will argue that no Bob-type measurements are possible. In the first place our example has $\exists_{x\in \mathbb{R}^2,|x|=1}\,B(x,\lambda) \not \equiv -A(x,\lambda)$ and shows critical violating parameter settings. It can be arranged such that Bob does not know $A$. Secondly, no local $\lambda_B$, observable by Bob in principle, is/are present in the evaluation of
\[
[a(t_0)b(t_1)]_{AA}\propto \int d\lambda_A \rho_A(\lambda_A)A(a(t_0),\lambda_A)A(b(t_1),\lambda_A)
\]
Bob will not know anything about the A measurements during the experiment. Alice and Bob can however beforehand agree on the pairwise settings $a(t_0)$ at Alice site and $b(t_1)$ at both Alice and Bob's site at $t=t_1$. At $t=t_2$ and $t=t_3$ the situation at $t_0$ and $t_1$ can be repeated etc.

If moreover, the entanglement correlation is considered valid only for $||A-S||_e=||S-B||_e$ experiments, then entanglement is narrowed down unwarrantedly. Hence, it is claimed here that (\ref{Dis1}) does not capture {\em all }possible models. Experiments based on (\ref{Dis1}) cannot rightfully claim to have covered {\em all } local and causal models. Interestingly, the numerical study presented here produces a (numerical) discrepancy, higher than two decimals precision, for $[ab]_{AA}$ and $E(a,b)$ when using violating pairs ( $\Delta=|\,|E(a,c)|-|[ac]_{AA}|\,|$ in  table-\ref{TB5}). 

It must be remarked that $E(b,b)=-1$ was not based on numerical considerations of the model. It is possible that the theoretical $E(b,b)=-1$ relation does not hold in a numerical sense in the model. Strictly speaking a discussion can then be started if one may use this relation.  Let us first note that denying $E(b,b)=-1$, disables local hidden variables models beforehand. Secondly, the argument in favour of the use of $E(b,b)=-1$ for the model, is presented in equation (\ref{Extr5})  and starts with the, obviously valid, theoretical derivation of $[ab]_{AA}=-E(a,b)$. Theoretically one cannot deny $E(b,b)=-1$ but it needs to be checked for the model. In order to do that let us substitute $a=b$ in $[ab]_{AA}=-E(a,b)$ and plug in the {\em model definition} $\mathcal{A}_1(\phi_b)=\sgn(A_0(\phi_b)-\lambda_A)$ in the formula for $[bb]_{AA}$. Obviously, $\mathcal{A}_1^2(\phi_b)=1$, hence, $[bb]_{AA}=1$ and so the model allows for $E(b,b)=-1$. This is sufficient and at the same time a necessary result. We note that a discussion on numerics vs derivation will show some traits of the Bouwer - Hilbert discussion \cite{3a} where construction versus arguing from inference is the bottom line. 

Furthermore, the objection that not a complete counter model is produced is not serious. The absence of a complete model is not safeguarding the inequalities from inconsistencies. Similar remarks can be made regarding a complete CHSH violating computer model. Moreover, \cite{3} rejects the validity of the CHSH. Because there is neither a direct experimental disprove nor a direct experimental prove of locality, it is reasonable to check all the assumptions in locality excluding tests. The paper, with sufficient argument, contests that Bell's method provides a simple and clear support for no-go locality and causality in quantum reality. It questions the support for definitions of a correlation approximated by heuristic counting measures presented as an obvious truth and reports results regarding the basis of Bell's correlation function itself. In order to answer the obvious objection: "what is wrong with Bell's approach", we refer to the possibility of concrete mathematical incompleteness \cite{4}, i.e. the impossibility to demonstrate truth or falsity of concrete mathematical claims from  basic axioms. Incompleteness in concrete mathematics is researchable in an exact manner \cite{5}. Given the Bell operations leading to inequalities and put those next to the arguments in the present paper, there is room for claiming that similar basic axioms in both cases were used. Nevertheless,  no definite final accept or reject conclusion can be arrived at. Hence, concrete incompleteness. Finally we also note that no positive proof for locality and causality is given either. It can be that the same logical linguisitic gap that lies behind mathematical incompleteness can be found behind locality and causality for quantum theory, i,e, Ludwig Wittgenstein's claim (number 5.1361) that causality is a form of superstition viz. Bertrand Russel's introduction to Wittgensteins Tractatus \cite{7}. Apparently does this superstition work reasonably well for the "classical" world but the researcher might cross a meaningful language border when causality is applied to the quantum domain. However, the author would like to refer the reader to work of Sanctuary \cite{8}. Perhaps that this "meaningful language" border is not a fixed line.


\begin{thebibliography}{10}
\bibitem{1} Bell J.S. { Physics} \textbf{1}, 195-200 (1964).
\bibitem{2} Einstein A., Podolsky B. and Rosen N., { Phys Rev} \textbf{47}, 777-780 (1935).
\bibitem{2a} Clauser J.F., Horne M.A., Shimony A. and Holt R.A. , {Phys Rev Lett} \textbf{23} 880-884 (1969).
\bibitem{3a} Brouwer, L.E.J., {\it Intuitionisme en formalisme}, Reprint of a lecture, P. Noordhof, Groningen The Netherlands, (1919).
\bibitem{3} Geurdes J.F. {Res in Phys} \textbf{4}, 81-82 (2014).
\bibitem{4} Friedman H.M. {\it Lecture notes}: Concrete Mathematical Incompleteness {Univ. Cambridge UK} (2010).
\bibitem{5} Friedman H.M. {\it Concrete mathematical incompleteness}, philpapers.org/rec/FRICMI, (2013).
\bibitem{7} Wittgenstein, L. {\em Tractatus Logico-Philosophicus } Logisch Philosopisches Abhandlung, KeganPaul (1922) or http://people.umass.edu/klement/tlp/.
\bibitem{8} Sanctuary, B.C., {\em The Dirac Equation in Two Dimensions}, submitted (2015).
\end{thebibliography}
\end{document}